\documentclass[namedreferences]{kluwer}

\usepackage{graphicx}
\begin{document}
\begin{article}

\begin{opening}
%<opening commands>

\title{Photometric Redshifts for an Optical/Near-Infrared Catalogue in the Chandra Deep Field South}

\author{Elizabeth R. \surname{Stanway}\thanks{ers@ast.cam.ac.uk}}
\author{Andrew \surname{Bunker}}
\author{Richard G. \surname{McMahon}}
\institute{Institute of Astronomy, Madingley Road, Cambridge, CB3 0HA}

%\dedication{...}
%\translation{...}

\runningtitle{Photometric Redshifts for a Catalogue in the CDFS}
\runningauthor{E R Stanway, A Bunker, R G McMahon}

%\begin{ao}
%....
%\end{ao}

\begin{abstract}
Photometric redshifts have proven a powerful tool in identifying galaxies over a large range of lookback times. We have been generalising this technique to incorporate the selection of candidate high redshift QSOs. We have applied this to a large optical/near-infrared imaging survey in 6 wavebands aiming to push farther in redshift (and fainter in luminosity) than previous studies. We believe that study of these very faint and distant objects provides valuable insights into galaxy formation and evolution.\\
Here we present work in progress and preliminary results for a catalogue of objects detected as part of the Las Campanas Infrared Survey. This is a stepping stone to the type of survey data that will become available in the next few years from projects such as UKIDSS and VISTA.\\

\end{abstract}

\end{opening}

%<body of paper>

\section{Theory and Method}

The technique of photometric redshifts has been extensively studied and developed in the last decade.  By comparing the flux observed from an object in a number of wavebands to that expected from templates of known  type, age and redshift, all these parameters can be found.  A publicly available application designed to perform $\chi^2$ minimisation over such a parameter space is Hyperz \cite{b2000}. This code has now been used and tested by several authors.\\
\\
Clearly, since objects are classified according to the best fitting template, the choice of which templates to provide has significant effects on the results obtained.  Hyperz is supplied together with a set of empirical galaxy templates \cite{cww} and a set of evolutionary synthesis galaxy templates \cite{bc}.\\
\\
We have been investigating the use of other template sets in Hyperz and in particular the inclusion of QSO and stellar templates.\\
\\
A model quasar spectral energy distribution (SED) has been developed from the Sloan Digital Sky Survey results and templates at a number of redshifts produced.  These model templates are then used as additional inputs for Hyperz.
\\
\\
The Star/Galaxy parameter in the SExtractor software has been used to divide catalogues into `extended' and `pointlike' objects.  A local QSO template may then be included with the galaxy SED set when studying extended objects.\\
\\
Pointlike objects are considered to be quasars or stars. For an initial analysis of these objects a new set of  templates are constrained to lie at z$=$0. This template set comprises a number of stellar templates from the Bruzual 77 library as well as an array of QSO templates at different redshifts.  Objects are then classified as stellar or QSO according to the best fitting template.
\\\\
The method has been tested on real and simulated data.  The reliability of the separation depends on the filter set being used but on simulated catalogues in {\it UBVRIH} we have found that fewer than 5\% of objects in the catalogue are misclassified in this way and fewer than 1\% of high redshift objects are assigned stellar templates or {\it vice versa}.\\

\section{Data, Tests and Results}

\begin{figure}
\begin{center}
\includegraphics[width=9.5cm, height=5.5cm]{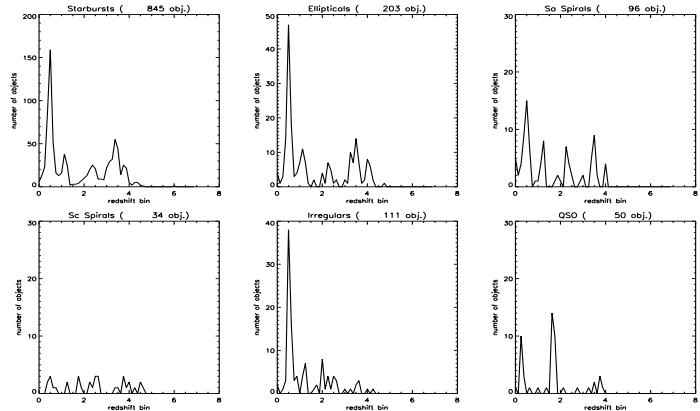}
\caption{ The Number Distribution of Galaxies for this {\it UBVRIH} catalogue.\label{results}}
\end{center}
\end{figure}

The data presented here were gathered as part of the Las Campanas Infrared Survey (LCIRS), a large survey covering an area of 1.2 deg$^2$ to faint magnitudes in 9 wavebands \cite{lcirs}. The survey is infrared driven and aims to improve understanding of galaxy evolution and large scale structure at $1<z<2$.  The data analysed here comprises a {\it VRIH} band photometric catalogue (with a limiting magnitude in H of 22.5) overlapping the Chandra Deep Field South (CDFS).  The redshift and nature of each object was considered and candidates for spectroscopic follow-up identified.\\
\\
Although photometric redshifts can be estimated from  just a few wavebands, their reliability may be poor due to degeneracies in colour-redshift space. A nearby age-reddened galaxy, for example, may have identical colours to a distant young galaxy. To  explore this problem several model catalogues were produced.  Photometric redshift estimates were compared to the model redshifts of  each object and the effects of excluding or including photometric bands explored.\\
\\
\noindent
As Figure \begin{figure}
\begin{center}
\includegraphics[width=8cm]{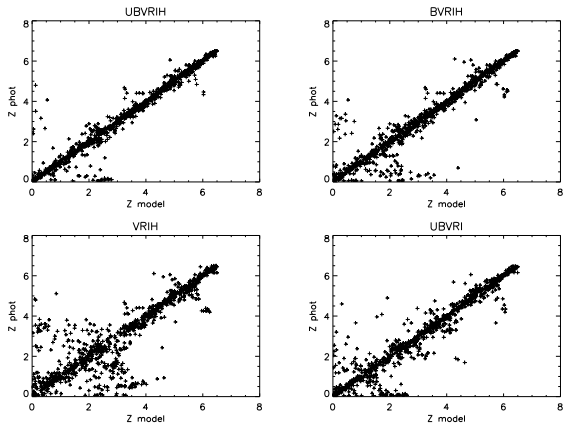}
\caption{ The performance of Hyperz on model catalogues with varying wavelength coverage. Note that the lack of either short or long wavelength bands significantly increases the chance of catastrophic failures.}\label{f1}
\end{center}
\end{figure} \ref{f1} shows Hyperz performs poorly when only the {\it VRIH} filters of LCIRS  are available.  The inclusion of {\it U} and  {\it B} band data drastically reduces the number of catastrophic failures for galaxies. For quasars on the other hand a colour degeneracy still remains in these bands.\\
\\
As a result of these and other tests, it was decided to obtain ESO imaging data for this field, publicly available as part of the ESO Imaging Project \cite{R98}. 
Although the ESO field is smaller than the LCIRS field, a subset of  1825 of the original 4135 objects with $H<22.5$ are in the ESO region.  For  these objects {\it U} and {\it B} band data has been combined with the LCIRS data to produce redshift estimates. The $\Delta(z)/(1+z)$ of these photometric redshifts is $\sim 0.09$ (from model catalogues).\\
\\
The results are presented in Figure \ref{results}. 
 Galaxy number counts  fall off strongly with redshift but a number of objects are assigned photometric redshifts $ > 4$. These candidate high-redshift galaxies would make promising targets for our intended spectroscopic follow-up.  In addition, we now have redshifts for a sample of Extremely Red Objects in this field (in collaboration with Richard Ellis, Pat McCarthy and Andrew Firth) which will be used to further calibrate the photometric redshift analysis.

\section{Conclusions}

This work is intended to be the basis for further study  both of the photometric redshift  technique and its applications to galaxy evolution.   We are now developing this field by  refining and applying semi-analytic galaxy models to better describe the universe at large lookback times.\\
\\
It is our belief that  galaxy redshift distributions obtained in this fashion from large optical-infrared surveys may provide valuable constraints on simulations and models of galaxy formation at large redshifts.\\

\end{article}

\begin{thebibliography}{}

\bibitem[\protect\citeauthoryear{}{}]{}

\bibitem[\protect\citeauthoryear{Bolzonella et al.}{2000}]{b2000}
Bolzonella, M., Miralles, J.-M. and Pell{\' o}, R.: 2000,
`Photometric redshifts based on standard SED fitting procedures', 
{\it A\&A\/} {\bf 363},
pp.~476--492

\bibitem[\protect\citeauthoryear{Coleman et al.}{1980}]{cww}
Coleman, G. D., Wu, C.-C. and Weedman, D. W.: 1980,
`Colors and magnitudes predicted for high redshift galaxies',
{\it ApJS\/} {\bf 43},
pp.~393--416

\bibitem[\protect\citeauthoryear{Bruzual and Charlot}{1993}]{bc}
Bruzual, A. G., and Charlot, S.: 1993,
`Spectral evolution of stellar populations using isochrone synthesis',
{\it AJ\/} {\bf 405},
pp.~538--553


\bibitem[\protect\citeauthoryear{Rengelink et al.}{1998}]{R98}
Rengelink, R., Nonino,M., da Costa, L., Zaggia, S., Erben, T., Benoist, C., Wicenec, A., Scodeggio, M., Olsen, L. F., Guarnieri, D., Deul, E., Hook, R., Moorwood, A. and Slijkhuis, R.: 1998,
`ESO Imaging Survey. AXAF Field',
{\it astro-ph\/}/9812190

\bibitem[\protect\citeauthoryear{McCarthy et al.}{2001}]{lcirs}
McCarthy, P. J., Carlberg, R. G., Chen, H.-W., Marzke, R. O., Firth, A. E., Ellis, R. S., Persson, S. E., McMahon, R. G., Lahav, O., Wilson, J., Martini, P., Abraham, R. G., Sabbey, C. N., Oemler, A., Murphy, D. C., Somerville, R. S., Beckett, M. G., Lewis, J. R., MacKay, C. D.: 2001,
`The Las Campanas Infrared Survey: Early-Type Galaxy Progenitors beyond z=1',
{\it AJ} {\bf 560}, 
pp.~L131--L134


\end{thebibliography}
\end{document}